\pdfoutput=1

\documentclass[aps,pra,twocolumn,twoside,preprintnumbers,showpacs,showkeys,superscriptaddress,byrevtex,floatfix,nofootinbib,amsmath,amssymb]{revtex4-1}

\usepackage{array,dsfont,graphicx,mathtools,multirow,subfig}
\usepackage[bookmarksnumbered,bookmarksopen,bookmarksopenlevel=2,pdfstartview=FitH]{hyperref}
\usepackage[all]{hypcap}

\newcolumntype{M}[1]{>{$}{#1}<{$}}
\newcolumntype{D}[1]{>{$\displaystyle}{#1}<{$}}

\DeclarePairedDelimiter{\bra}{\langle}{\rvert}
\DeclarePairedDelimiter{\ket}{\lvert}{\rangle}

\DeclareMathOperator{\tr}{tr}
\DeclareMathOperator{\Tr}{Tr}
\DeclareMathOperator{\Det}{Det}
\DeclareMathOperator{\Der}{Der}
\DeclareMathOperator{\Aut}{Aut}
\DeclareMathOperator{\Str}{Str}

\newcommand{\half}{\tfrac{1}{2}}
\newcommand{\rep}[1]{\mathbf{#1}}
\newcommand{\braket}[1]{\langle#1|#1\rangle}
\newcommand{\AutM}{\Aut(\mathfrak{M}(\mathfrak{J}_\mathds{C}))}

\begin{document}

\title{Freudenthal triple classification of three-qubit entanglement}
\author{L. Borsten}
\email[]{leron.borsten@imperial.ac.uk}
\author{D. Dahanayake}
\email[]{duminda.dahanayake@imperial.ac.uk}
\author{M. J. Duff}
\email[]{m.duff@imperial.ac.uk}
\affiliation{Theoretical Physics, Blackett Laboratory, Imperial College London, London SW7 2AZ, United Kingdom}
\author{H. Ebrahim}
\email[]{hebrahim@brandeis.edu}
\affiliation{Theory Group, Martin Fisher School of Physics, Brandeis University, MS057, 415 South St., Waltham, MA 02454, U.S.A.}
\author{W. Rubens}
\email[]{william.rubens06@imperial.ac.uk}
\affiliation{Theoretical Physics, Blackett Laboratory, Imperial College London, London SW7 2AZ, United Kingdom}
\setcounter{affil}{1}
\date{\today}

\begin{abstract}

We show that the three-qubit entanglement classes: (0) Null, (1) Separable $A$-$B$-$C$, (2a) Biseparable $A$-$BC$, (2b) Biseparable $B$-$CA$, (2c) Biseparable $C$-$AB$, (3) W and (4) GHZ correspond respectively to ranks 0, 1, 2a, 2b, 2c, 3 and 4 of a Freudenthal triple system defined over the Jordan algebra $\mathds{C\oplus C\oplus C}$. We also compute the corresponding SLOCC orbits.

\end{abstract}



\preprint{Imperial/TP/2008/mjd/4}
\preprint{BRX-TH 605}

\maketitle

\section{Introduction}
\label{sec:introduction}

Quantum entanglement lies at the heart of quantum information theory, with applications to quantum computing, teleportation, cryptography and communication \cite{Nielsen:2000}. The case of three qubits (Alice, Bob, Charlie) is particularly interesting \cite{Linden:1997qd,Kempe:1999vk,Dur:2000,Acin:2000,Carteret:2000-1,Sudbery:2001,Carteret:2000-2,Levay:2004,Brody:2007} since it provides the simplest example of inequivalently entangled states. It is by now well understood that  there are seven entanglement classes: (0) Null, (1) Separable $A$-$B$-$C$, (2a) Biseparable $A$-$BC$, (2b) Biseparable $B$-$CA$, (2c) Biseparable $C$-$AB$, (3) W and (4) GHZ.   We summarise this conventional  classification of three-qubit entanglement in \autoref{sec:conventional}.

The purpose of the present paper is to give a novel version of this classification by invoking that elegant branch of mathematics  involving Jordan algebras and Freudenthal triple systems (FTS).  In particular we note that an FTS is characterised by its rank: 0 to 4. (The relevant mathematics is  briefly reviewed in \hyperref[sec:JordanandFTS]{appendix A}).

By making the following direct correspondence between a three-qubit state vector $\ket{\psi}$ and a Freudenthal triple system $\Psi$ over the Jordan algebra $\mathds{C\oplus C\oplus C}$:
\begin{equation}
\begin{split}
\ket{\psi}&=a_{ABC}\ket{ABC}\\\leftrightarrow\quad\Psi&=\begin{pmatrix}a_{111}&(a_{001}, a_{010}, a_{100})\\(a_{110}, a_{101}, a_{011})&a_{000}\end{pmatrix},
\end{split}
\label{eq:3bit}
\end{equation}
we show in \autoref{sec:classification}  that the structure of the FTS naturally captures the Stochastic Local Operations and Classical Communication (SLOCC) classification described in \autoref{sec:conventional}.  The entanglement classes correspond to FTS ranks 0, 1, 2a, 2b, 2c, 3 and 4, respectively.

This also facilitates a computation of the SLOCC orbits.

\section{Conventional three-qubit entanglement classification}
\label{sec:conventional}

The concept of entanglement is the single most important feature distinguishing classical information theory from quantum information theory. We may naturally describe and harness entanglement by the protocol of Local Operations and Classical Communication (LOCC).  LOCC describes a multi-step process for transforming any input state to a different output state while obeying certain rules.  Given any multipartite state, we may split it up into its relevant parts and send each of them to different labs around the world. We allow the respective scientists to perform any experiment they see fit; they may then communicate these results to each other classically (using email or phone or carrier pigeon). Furthermore, for the most general LOCC, we allow them to do this as many times as they like.  Any classical correlation may be experimentally established using LOCC. Conversely, all correlations not achievable via LOCC are attributed to genuine quantum correlations.

Since  LOCC cannot create entanglement, any two states which may be interrelated using LOCC ought to be physically equivalent with respect to their entanglement properties.  Two states of a composite system are LOCC equivalent if and only if they may be transformed into one another using the group of \emph{local unitaries} (LU), unitary transformations which factorise into separate transformations on the component parts  \cite{Bennett:1999} .   In the case of $n$ qudits, the LU group (up to a phase) is given by $\left[SU(d)\right]^n$.  For unnormalised three-qubit states, the number of parameters \cite{Linden:1997qd} needed to describe inequivalent states or, what amounts to the same thing, the number of algebraically independent invariants \cite{Sudbery:2001} is thus given by the dimension of the space of orbits
\begin{equation}
\frac{\mathds{C}^2\times\mathds{C}^2\times\mathds{C}^2}{U(1) \times SU(2) \times SU(2) \times SU(2)},
\end{equation}
namely $16-10=6$.  These six invariants  are given as follows.

\begin{description}
\item[1] The norm squared:
\begin{equation}
|\psi|^2=\braket{\psi}.
\end{equation}
\item[2A, 2B, 2C] The local entropies:
 \begin{equation}
\begin{split}
S_{A}&=4\det\rho_A, \\
S_{B}&=4\det\rho_B, \\
S_{C}&=4\det\rho_C,
\end{split}
\end{equation}
where $\rho_A, \rho_B, \rho_C$ are the doubly  reduced density matrices:
\begin{equation}
\begin{split}
\rho_A&=\Tr_{BC}\ket{\psi}\bra{\psi}, \\
\rho_B&=\Tr_{CA}\ket{\psi}\bra{\psi}, \\
\rho_C&=\Tr_{AB}\ket{\psi}\bra{\psi}.
\end{split}
\end{equation}
\item[3] The Kempe invariant \cite{Kempe:1999vk,Sudbery:2001, Osterloh:2008,Osterloh:2008b}:
\begin{equation}
\begin{split}
K&=\tr(\rho_A\otimes\rho_B\rho_{AB})-\tr(\rho_{A}^{3})-\tr(\rho_{B}^{3}) \\
&=\tr(\rho_B\otimes\rho_C\rho_{BC})-\tr(\rho_{B}^{3})-\tr(\rho_{C}^{3}) \\
&=\tr(\rho_C\otimes\rho_A\rho_{CA})-\tr(\rho_{C}^{3})-\tr(\rho_{A}^{3}),
\end{split}
\end{equation}
where $\rho_{AB}, \rho_{BC}, \rho_{CA}$ are the singly  reduced density matrices:
\begin{equation}
\begin{split}
\rho_{AB}&=\Tr_{C}\ket{\psi}\bra{\psi}, \\
\rho_{BC}&=\Tr_{A}\ket{\psi}\bra{\psi}, \\
\rho_{CA}&=\Tr_{B}\ket{\psi}\bra{\psi}.
\end{split}
\end{equation}

\item[4] The 3-tangle \cite{Coffman:1999jd}
\begin{equation}
\tau_{ABC}=4|\Det a_{ABC}|
\end{equation}
where $a_{ABC}$ are the state coefficients appearing in \eqref{eq:3bit} and where $\Det a_{ABC}$ is Cayley's hyperdeterminant \cite{Cayley:1845,Miyake:2002}:
\begin{equation}\label{eq:hyperdet}
\begin{gathered}
\Det a_{ABC}:=\\
\begin{aligned}
&-\half~\varepsilon^{A_1A_2}\varepsilon^{B_1B_2}\varepsilon^{A_3A_4}\varepsilon^{B_3B_4}\varepsilon^{C_1C_4}\varepsilon^{C_2C_3}\\ &\phantom{\quad}\times a_{A_1B_1C_1}a_{A_2B_2C_2}a_{A_3B_3C_3}a_{A_4B_4C_4}.
\end{aligned}
\end{gathered}
\end{equation}
Here $\varepsilon$ is the $SL(2, \mathds{C})$--invariant alternating tensor
\begin{equation}\label{eq:ospgroupdef}
\begin{gathered}
\varepsilon:=\begin{pmatrix}0&1\\-1&0\end{pmatrix},
\end{gathered}
\end{equation}
We also adopt the Einstein summation convention that repeated indices are summed over.
\end{description}

The LU orbits partition the Hilbert space into equivalence classes.
However, for single copies of pure states this classification is both mathematically and physically too restrictive. Under LU two states of even the simplest bipartite systems will not, in general, be related \cite{Dur:2000}. Continuous parameters are required to describe the space of entanglement classes \cite{Linden:1997qd,Carteret:2000-1,Sudbery:2001,Carteret:2000-2}. In this sense the LU classification is too severe \cite{Dur:2000}, obscuring some of the more qualitative features of entanglement. An alternative classification scheme was proposed in \cite{Bennett:1999, Dur:2000}. Rather than declare equivalence when states are deterministically related to each other by LOCC, we require only that they may be transformed into one another with  some non-zero probability of success.

This coarse graining goes by the name of Stochastic LOCC or SLOCC for short. Stochastic LOCC includes, in addition to LOCC, those quantum operations that are not trace-preserving on the density matrix,  so that we no longer require that the protocol always succeeds with certainty. It is proved in \cite{Dur:2000} that for $n$ qudits, the SLOCC equivalence group is (up to an overall complex factor) $\left[SL(d, \mathds{C})\right]^n$. Essentially, we may identify two states if there is a non-zero probability that one can be converted into the other and vice-versa, which means we get $[SL(d, \mathds{C})]^n$ orbits rather than the $[SU(d)]^n$ kind of LOCC. This generalisation may be physically motivated by the fact that any set of SLOCC equivalent states may be used to perform the same non-classical operations, only with varying likelihoods of success.

In the case of three qubits, the  group of invertible SLOCC transformations is $SL(2,\mathds{C}) \times SL(2,\mathds{C}) \times SL(2,\mathds{C})$.  Tensors transforming under the Alice, Bob or Charlie $SL(2,\mathds{C})$ carry indices $A_1,A_2...$, $B_1,B_2...$ or $C_1,C_2...$, respectively, so $a_{ABC}$ transforms as a $\rep{(2,2,2)}$. Hence the hyperdeterminant \eqref{eq:hyperdet} is manifestly SLOCC invariant.  Further, under this coarser SLOCC classification, D\"ur et al. \cite{Dur:2000} used simple arguments concerning the conservation of ranks of reduced density matrices to show that there are only six three-qubit equivalence classes (or seven if we count the null state); only two of which show
\emph{genuine} tripartite entanglement.  They are as follows.
\captionsetup{justification=raggedright}
\begin{figure}
\centering
\subfloat[Onion structure]{\label{fig:orbits}\includegraphics[width=.9\linewidth]{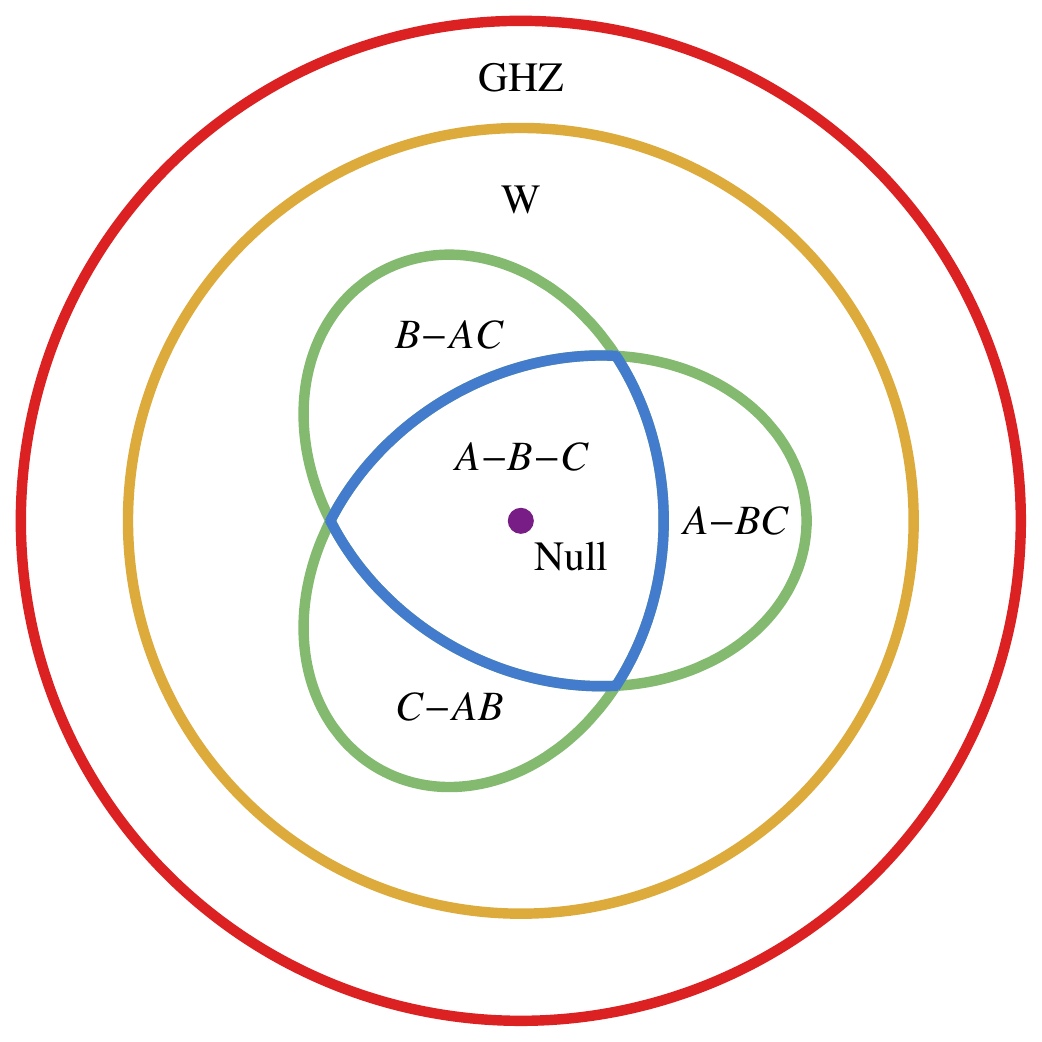}}\\
\subfloat[Hierarchy]{\label{fig:hierarchy}\includegraphics[width=\linewidth]{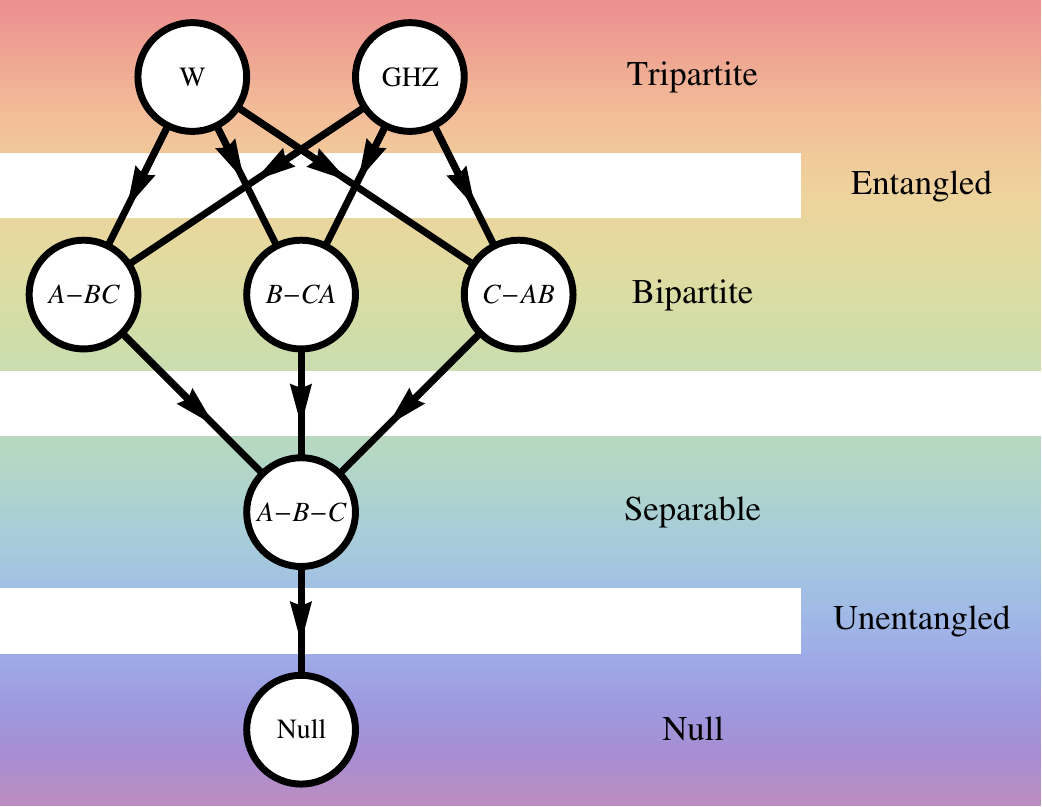}}
\caption{(a) Onion-like classification of SLOCC orbits. (b) Stratification. The arrows are non-invertible SLOCC transformations between classes that generate the entanglement hierarchy. The partial order defined by the arrows is transitive, so we may omit e.g. GHZ $\to$ $A$-$B$-$C$ and $A$-$BC$ $\to$ Null arrows for clarity.\label{fig:conventional}}
\end{figure}
\begin{table}
\caption{The values of the local entropies $S_A, S_B$, and $S_C$ and the hyperdeterminant $\Det a$ are used to partition three-qubit states into entanglement classes.\label{tab:conventional}}
\begin{ruledtabular}
\begin{tabular}{cc*{8}{M{c}}c}
& \multirow{2}{*}{Class} & \multirow{2}{*}{Representative} & & \multicolumn{5}{c}{Condition}          & & \\
\cline{4-10}
&                        &                                 & & \psi  & S_A   & S_B   & S_C   & \Det a & & \\
\hline
& Null                   & 0                               & & =0    & =0    & =0    & =0    & =0     & & \\
& $A$-$B$-$C$            & \ket{000}                       & & \neq0 & =0    & =0    & =0    & =0     & & \\
& $A$-$BC$               & \ket{010}+\ket{001}             & & \neq0 & =0    & \neq0 & \neq0 & =0     & & \\
& $B$-$CA$               & \ket{100}+\ket{001}             & & \neq0 & \neq0 & =0    & \neq0 & =0     & & \\
& $C$-$AB$               & \ket{010}+\ket{100}             & & \neq0 & \neq0 & \neq0 & =0    & =0     & & \\
& W                      & \ket{100}+\ket{010}+\ket{001}   & & \neq0 & \neq0 & \neq0 & \neq0 & =0     & & \\
& GHZ                    & \ket{000}+\ket{111}             & & \neq0 & \neq0 & \neq0 & \neq0 & \neq0  & &
\end{tabular}
\end{ruledtabular}
\end{table}
\begin{description}
  \item[Null:] The trivial zero entanglement orbit corresponding to vanishing states,
      \begin{equation}\text{Null}:\quad0.\end{equation}
  \item[Separable:] Another zero entanglement orbit for completely factorisable product states,
      \begin{equation}A\text{-}B\text{-}C:\quad\ket{000}.\end{equation}
  \item[Biseparable:] Three classes of bipartite entanglement
      \begin{equation}
      \begin{split}
      A\text{-}BC:\quad\ket{010}+\ket{001},\\
      B\text{-}CA:\quad\ket{100}+\ket{001},\\
      C\text{-}AB:\quad\ket{010}+\ket{100}.
      \end{split}
      \end{equation}
  \item[W:] Three-way entangled states that do not maximally violate Bell-type inequalities in the same way as the GHZ class discussed below. However, they are robust in the sense that tracing out a subsystem generically results in a bipartite mixed state that is maximally entangled under a number of criteria \cite{Dur:2000},
      \begin{equation}\text{W}:\quad\ket{100}+\ket{010}+\ket{001}.\end{equation}
  \item[GHZ:] Genuinely tripartite entangled Greenberger-Horne-Zeilinger \cite{Greenberger:1989} states. These maximally violate Bell-type inequalities but, in contrast to class W, are fragile under the tracing out of a subsystem since the resultant state is completely unentangled,
      \begin{equation}\text{GHZ}:\quad\ket{000}+\ket{111}.\end{equation}
\end{description}
These classes and the above representative states from each class are summarised in \autoref{tab:conventional}. They are characterised \cite{Dur:2000} by the vanishing or not of the invariants listed in the table. Note that the Kempe invariant is redundant in this SLOCC classification. A visual representation of these SLOCC orbits is provided by the onion-like classification \cite{Miyake:2002} of \hyperref[fig:orbits]{Figure 1a}.

These SLOCC equivalence classes are then stratified by \emph{non-invertible} SLOCC operations into an entanglement hierarchy \cite{Dur:2000} as depicted in \hyperref[fig:hierarchy]{Figure 1b}. Note that no SLOCC  operations (invertible or not) relate the GHZ and W classes; they are genuinely distinct classes of tripartite entanglement. However, from either the GHZ class or W class one may use non-invertible SLOCC transformations to descend to one of the biseparable or separable classes and hence we have a hierarchical entanglement structure.

\section{The FTS classification of qubit entanglement}
\label{sec:classification}

\subsection{FTS representation of three-qubits}
\label{FTSthreequbits}

The goal of this section is to show that the classification of three qubits can be replicated in the completely different mathematical language of Jordan algebras and Freudenthal triple systems. A Jordan algebra $\mathfrak{J}$ is vector space defined over a ground field $\mathds{F}$ equipped with a bilinear product satisfying
\begin{equation}\label{eq:Jid}
\begin{split}
A\circ B &=B\circ A,\\ A^2\circ (A\circ B)&=A\circ (A^2\circ B), \quad\forall\ A, B \in \mathfrak{J}.
\end{split}
\end{equation}
One is then able to construct an FTS by defining the vector space $\mathfrak{M(J)}$,
\begin{equation}
\mathfrak{M(J)}=\mathds{F\oplus F}\oplus \mathfrak{J\oplus J}.
\end{equation}
An arbitrary element $x\in \mathfrak{M(J)}$ may be written as a ``$2\times 2$ matrix'',
\begin{equation}
x=\begin{pmatrix}\alpha&A\\B&\beta\end{pmatrix} \quad\text{where} ~\alpha, \beta\in\mathds{F}\quad\text{and}\quad A, B\in \mathfrak{J}.
\end{equation}
The relevant details of these constructions are spelled out in \hyperref[sec:JordanandFTS]{appendix A}. The FTS comes equipped with a quadratic form $\{x, y\}$,  a triple product $T(x,y,z)$ and a quartic norm $q(x, y, w, z)$, as defined in \eqref{eq:bilinearform}, \eqref{eq:tripleproduct} and \eqref{eq:quarticnorm}. Of particular importance is the \emph{automorphism group} $\Aut(\mathfrak{M(J)})$ given by the set of all transformations which leave invariant both the quadratic form and the quartic norm $q(x, y, w, z)$  \cite{Brown:1969}.

Following \cite{Krutelevich:2004},  the Jordan algebras, the Freudenthal triple systems,   and their associated automorphism groups, are summarised in \autoref{tab:FTSsummary}.  The conventional concept of matrix rank may be generalised to Freudenthal triple systems in a natural and $\Aut(\mathfrak{M(J)})$ invariant manner. The rank of an arbitrary element $x\in\mathfrak{M(J)}$ is uniquely defined using the relations in \autoref{tab:FTSrank} \cite{Ferrar:1972, Krutelevich:2004}.
\begin{table}
\caption{The Lie group and the dimension of its representation given by the Freudenthal construction defined over the cubic Jordan algebra $\mathfrak{J}$. The case $\mathfrak{J} = \mathds{F\oplus F\oplus F}$ with $\mathds{F=C}$ will be the FTS used to represent three qubits.\label{tab:FTSsummary}}
\begin{ruledtabular}
\begin{tabular}{*{6}{M{c}}}
& \text{Jordan algebra\ }\mathfrak{J}        & \dim\mathfrak{J} & \Aut(\mathfrak{M(J)})         & \dim\mathfrak{M(J)} & \\
\hline
& \mathds{F}                                 & 1                & SL(2)                         & 4                   & \\
& \mathds{F}\oplus\mathds{F}                 & 2                & SL(2)\times SL(2)			    & 6                   & \\
& \mathds{F}\oplus\mathds{F}\oplus\mathds{F} & 3                & SL(2)\times SL(2)\times SL(2) & 8                   & \\
& J_{3}^{\mathds{R}}                         & 6                & C_3                           & 14                  & \\
& J_{3}^{\mathds{C}}                         & 9                & A_5                           & 20                  & \\
& J_{3}^{\mathds{H}}                         & 15               & D_6                           & 32                  & \\
& J_{3}^{\mathds{O}}                         & 27               & E_7                           & 56                  & \\
& \mathds{F}\oplus Q_n                       & n+1              & SL(2)\times SO(n+2)           & 2n+4                &
\end{tabular}
\end{ruledtabular}
\end{table}

Our FTS representation of three qubits corresponds to the special case of \autoref{tab:FTSsummary} where the Jordan algebra is simply  $\mathfrak{J}_\mathds{C}=\mathds{C\oplus C\oplus C}$.  Define the cubic form
\begin{equation}
\begin{split}
N(A)&=A_1A_2A_3\\
\end{split}
\end{equation}
where $A=(A_1, A_2, A_3)\in \mathfrak{J}_\mathds{C}$.  One finds, using \eqref{eq:cubicdefs},
\begin{equation}
\Tr(A,B)= A_1B_1+A_2B_2+A_3B_3,
\end{equation}
Then, using $\Tr(A^\sharp,B)=3N(A,A,B)$, the quadratic adjoint is given by
\begin{gather}
A^\sharp=(A_2A_3, A_1A_3, A_1A_2),\\
\shortintertext{and therefore}
\begin{split}
(A^\sharp)^\sharp&=(A_1A_2A_3A_1,A_1A_2A_3A_2,A_1A_2A_3A_3)\\
                 &=N(A)A.
\end{split}
\end{gather}
It is not hard to check $\Tr(A,B)$ is non-degenerate and so $N$ is Jordan cubic as described in \hyperref[sec:Jordan]{appendix A1}. Hence, we have a cubic Jordan algebra $\mathfrak{J}_\mathds{C}=\mathds{C\oplus C\oplus C}$ with product given by
\begin{equation}
\begin{split}
A\circ B&=(A_1B_1, A_2B_2, A_3B_3).
\end{split}
\end{equation}
The structure and reduced structure groups are given by $[SO(2, \mathds{C})]^3$ and $[SO(2, \mathds{C})]^2$ respectively.
\begin{table}
\caption{Partition of the space $\mathfrak{M(J)}$ into five orbits of $\Aut(\mathfrak{M(J)})$ or ranks.\label{tab:FTSrank}}
\begin{ruledtabular}
\begin{tabular}{ccc*{4}{M{c}}cc}
& \multirow{2}{*}{Rank} & & \multicolumn{4}{c}{Condition}                 & & \\
\cline{3-8}
&                       & & x     & 3T(x,x,y)+\{x,y\}x & T(x,x,x) & q(x)  & & \\
\hline
& 0                     & & =0    & =0\ \forall\ y     & =0       & =0    & & \\
& 1                     & & \neq0 & =0\ \forall\ y     & =0       & =0    & & \\
& 2                     & & \neq0 & \neq0              & =0       & =0    & & \\
& 3                     & & \neq0 & \neq0              & \neq0    & =0    & & \\
& 4                     & & \neq0 & \neq0              & \neq0    & \neq0 & &
\end{tabular}
\end{ruledtabular}
\end{table}

We are now in a position to employ the FTS $\mathfrak{M}(\mathfrak{J}_\mathds{C})=\mathds{C}\oplus \mathds{C}\oplus \mathfrak{J}_\mathds{C}\oplus\mathfrak{J}_\mathds{C}$ as the representation space of three qubits.  In this case, an element of the FTS is  given by
\begin{equation}\label{eq:BasicFts}
        \begin{pmatrix} \alpha & (A_1, A_2, A_3) \\
                        (B_1,B_2,B_3)   &  \beta
        \end{pmatrix}
\end{equation}
where $\alpha,\beta, A_1, A_2,A_3,  B_1, B_2, B_3 \in \mathds{C}$.
The essential purpose of this paper is to identify these eight complex numbers with the eight complex components of the three qubit wavefunction $\ket{\Psi} = a_{ABC}\ket{ABC}$,
\begin{equation}
\begin{split}
        &\begin{pmatrix} \alpha & (A_1, A_2, A_3) \\
                        (B_1,B_2,B_3)   &  \beta
        \end{pmatrix}\\ \leftrightarrow
       & \begin{pmatrix} a_{111} & (a_{001}, a_{010}, a_{100}) \\
                       (a_{110}, a_{101}, a_{011})&a_{000}
        \end{pmatrix}
\end{split}
\end{equation}
so that all the powerful machinery of the Freudenthal triple system may now be applied to qubits.

Using \eqref{eq:quarticnorm} one finds that the quartic norm $q(\Psi)$ is related to Cayley's hyperdeterminant by
\begin{equation}\label{eq:q}
\begin{split}
q(\Psi)&=\{T(\Psi,\Psi,\Psi),\Psi\}\\
       &=2\det\gamma^{A}=2\det\gamma^{B}=2\det\gamma^{C}\\
       &=-2\Det a_{ABC},
\end{split}
\end{equation}
where, following \cite{Duff:2006ev,Toumazet:2006,Borsten:2008wd} we have defined the three matrices $\gamma^A,\gamma^B$, and $\gamma^C$
\begin{gather}
\begin{split}\label{eq:ABCgammas}
(\gamma^{A})_{A_{1}A_{2}}&=\varepsilon^{B_{1}B_{2}}\varepsilon^{C_{1}C_{2}}a_{A_{1}B_{1}C_{1}}a_{A_{2}B_{2}C_{2}}, \\
(\gamma^{B})_{B_{1}B_{2}}&=\varepsilon^{C_{1}C_{2}}\varepsilon^{A_{1}A_{2}}a_{A_{1}B_{1}C_{1}}a_{A_{2}B_{2}C_{2}}, \\
(\gamma^{C})_{C_{1}C_{2}}&=\varepsilon^{A_{1}A_{2}}\varepsilon^{B_{1}B_{2}}a_{A_{1}B_{1}C_{1}}a_{A_{2}B_{2}C_{2}}.
\end{split}
\end{gather}
transforming respectively as $\rep{(3,1,1), (1,3,1), (1,1,3)}$  under $SL(2,\mathds{C}) \times SL(2,\mathds{C}) \times SL(2,\mathds{C})$. Explicitly,
\begin{widetext}
\begin{gather}
\begin{split}
\gamma^{A}&=
\begin{pmatrix}
2(a_{0}a_{3}-a_{1}a_{2}) &  a_{0}a_{7}-a_{1}a_{6}+a_{4}a_{3}-a_{5}a_{2}\\
a_{0}a_{7}-a_{1}a_{6}+a_{4}a_{3}-a_{5}a_{2}  & 2(a_{4}a_{7}-a_{5}a_{6})
\end{pmatrix}, \\
\gamma^{B}&=
\begin{pmatrix}2(a_{0}a_{5}-a_{4}a_{1}) & a_{0}a_{7}-a_{4}a_{3}+a_{2}a_{5}-a_{6}a_{1}\\
a_{0}a_{7}-a_{4}a_{3}+a_{2}a_{5}-a_{6}a_{1} & 2(a_{2}a_{7}-a_{6}a_{3})
\end{pmatrix}, \\
\gamma^{C}&=
\begin{pmatrix}
2(a_{0}a_{6}-a_{2}a_{4}) &  a_{0}a_{7}-a_{2}a_{5}+a_{1}a_{6}-a_{3}a_{4}\\
a_{0}a_{7}-a_{2}a_{5}+a_{1}a_{6}-a_{3}a_{4} & 2(a_{1}a_{7}-a_{3}a_{5})
\end{pmatrix},
\end{split}
\end{gather}
\end{widetext}
where we have made the decimal-binary conversion 0, 1, 2, 3, 4, 5, 6, 7 for 000, 001, 010, 011, 100, 101, 110, 111.
The $\gamma$'s are related to the local entropies of \autoref{sec:conventional} by
\begin{gather}
S_A = 4\Big[\tr\gamma^{B\dag}\gamma^B+\tr\gamma^{C\dag}\gamma^C\Big]\label{eq:S_i(gamma)}, \\
\tr\gamma^{A\dag}\gamma^A = \tfrac{1}{8}\big[S_B+S_C-S_A\big]\label{eq:gamma(S_i)}
\end{gather}
and their cyclic permutations.

The triple product maps a state $\Psi$, which transforms as a $\rep{(2,2,2)}$ of  $[SL(2, \mathds{C})]^3$, to another state $T(\Psi, \Psi, \Psi)$, cubic in the state vector coefficients, also transforming as a $\rep{(2,2,2)}$. Explicitly, $T(\Psi, \Psi, \Psi)$ may be written as
\begin{equation}\label{eq:Tdef}
T(\Psi, \Psi, \Psi)=T_{ABC}\ket{ABC}
\end{equation}
where $T_{ABC}$ takes one of three equivalent forms
\begin{equation}\label{eq:Tofgamma}
\begin{split}
T_{A_3B_1C_1}=\varepsilon^{A_1A_2}a_{A_1B_1C_1}(\gamma^{A})_{A_{2}A_{3}}\\
T_{A_1B_3C_1}=\varepsilon^{B_1B_2}a_{A_1B_1C_1}(\gamma^{B})_{B_{2}B_{3}}\\
T_{A_1B_1C_3}=\varepsilon^{C_1C_2}a_{A_1B_1C_1}(\gamma^{C})_{C_{2}C_{3}}.
\end{split}
\end{equation}
This definition permits us to link $T$ to the norm, local entropies and  the Kempe invariant  of \autoref{sec:conventional}:
\begin{equation}
\braket{T}=\tfrac{2}{3}(K-|\psi|^6)+\tfrac{1}{16}|\psi|^2(S_A+S_B+S_C).
\end{equation}
Having couched the three-qubit system within the FTS framework we may assign an abstract FTS rank to an arbitrary state $\Psi$
as in \autoref{tab:FTSrank}.

Strictly speaking, the automorphism group $\Aut(\mathfrak{M(J)})$ is not simply  $SL(2,\mathds{C}) \times SL(2,\mathds{C}) \times SL(2,\mathds{C})$ but includes a semi-direct product with the interchange triality $A \leftrightarrow B   \leftrightarrow C$.  The rank conditions of \autoref{tab:FTSrank} are invariant under this triality.  However, as we shall demonstrate, the set of rank 2 states may be subdivided into three distinct classes  which are inter-related by this triality. In the next section we show that these rank conditions give the correct entanglement classification of three qubits as in \autoref{tab:merge}.
\begin{table}
\caption{The entanglement classification of three qubits as according to the FTS rank system.\label{tab:merge}}
\begin{ruledtabular}
\begin{tabular}{ccccM{c}M{c}cc}
& \multirow{2}{*}{Class} & \multirow{2}{*}{Rank} & & \multicolumn{2}{c}{\text{FTS rank condition}}               & & \\
\cline{4-7}
&                        &                       & & \text{vanishing}                     & \text{non-vanishing} & & \\
\hline
& Null                   & 0  		             & & \Psi  				                  & -                    & & \\
& $A$-$B$-$C$            & 1  		             & & 3T(\Psi,\Psi,\Phi)+\{\Psi,\Phi\}\Psi & \Psi 		         & & \\
& $A$-$BC$               & 2a 		             & & T(\Psi,\Psi,\Psi) 					  & \gamma^A             & & \\
& $B$-$CA$               & 2b 		             & & T(\Psi,\Psi,\Psi) 					  & \gamma^B             & & \\
& $C$-$AB$               & 2c 		             & & T(\Psi,\Psi,\Psi)                    & \gamma^C	         & & \\
& W                      & 3  		             & & q(\Psi)							  & T(\Psi,\Psi,\Psi) 	 & & \\
& GHZ                    & 4  		             & & -                                    & q(\Psi)              & &
\end{tabular}
\end{ruledtabular}
\end{table}

\subsection{The FTS rank entanglement classes}
\label{sec:rank}

Rank 0 trivially corresponds to the vanishing state as in \autoref{tab:merge}. Since this implies vanishing norm, it is usually omitted from the entanglement discussion.

\subsubsection{Rank 1 and the class of separable states}
\label{sec:rank1}

A non-zero state $\Psi$  is rank 1 if
\begin{equation}\label{eq:FTSrank1}
\Upsilon :=3T(\Psi, \Psi, \Phi) + \{\Psi , \Phi\}\Psi=0, \quad \forall\ \Phi
\end{equation}
which implies, in particular,
\begin{equation}\label{eq:FTSrank2}
T(\Psi,\Psi,\Psi)=0.
\end{equation}
For the case $\mathfrak{J}_\mathds{C}=\mathds{C\oplus C\oplus C}$,
\begin{equation}
\begin{split}
(\gamma^A)_{A_{1}A_{2}}(\gamma^C)_{C_{1}C_{2}}&=\phantom{\times}\varepsilon^{B_1B_2}\varepsilon^{Z_1Z_2}\\
&\phantom{=}\times a_{A_1B_1Z_1}a_{A_2B_2Z_2}(\gamma^C)_{C_{1}C_{2}}\\
&=\phantom{+}\varepsilon^{B_2B_1}a_{A_1B_1C_1}T_{A_2B_2C_2}\\
&\phantom{=}+\varepsilon^{B_1B_2}a_{A_2B_2C_1}T_{A_1B_1C_2},
\end{split}
\end{equation}
and similarly for $(\gamma^B)_{B_{1}B_{2}}(\gamma^A)_{A_{1}A_{2}}$ and $(\gamma^C)_{C_{1}C_{2}}(\gamma^B)_{B_{1}B_{2}}$. So the weaker condition \eqref{eq:FTSrank2} means that at most only one of the gammas is non-vanishing.
From \eqref{eq:q}, moreover, it has vanishing determinant.  Furthermore,
\begin{gather}
\begin{split}
\Upsilon_{A_3B_1C_1}&=\phantom{\times}\varepsilon^{A_1A_2}\varepsilon^{B_2B_3}\varepsilon^{C_2C_3}\\
&\phantom{=}\times[\phantom{+\ }a_{A_1B_1C_1}a_{A_2B_2C_2}b_{A_3B_3C_3}\\
&\phantom{=\times\big[}+a_{A_1B_1C_1}b_{A_2B_2C_2}a_{A_3B_3C_3}\\
&\phantom{=\times\big[}+b_{A_1B_1C_1}a_{A_2B_2C_2}a_{A_3B_3C_3}\\
&\phantom{=\times\big[}-a_{A_1B_2C_2}b_{A_2B_3C_3}a_{A_3B_1C_1}]
\end{split}\\
\shortintertext{or}
\begin{split}
-\Upsilon_{A_1B_1C_1}&=\phantom{+}\varepsilon^{A_2A_3}b_{A_3B_1C_1}(\gamma^A)_{A_1A_2}\\
&\phantom{=}+\varepsilon^{B_2B_3}b_{A_1B_3C_1}(\gamma^B)_{B_1B_2}\\
&\phantom{=}+\varepsilon^{C_2C_3}b_{A_1B_1C_3}(\gamma^C)_{C_1C_2}
\end{split}\\
\shortintertext{where}
\begin{split}
\ket{\phi}&=b_{ABC}\ket{ABC}\\\leftrightarrow\quad\Phi&=\begin{pmatrix}b_{111}&(b_{001}, b_{010}, b_{100})\\(b_{110}, b_{101}, b_{011})&b_{000}\end{pmatrix}.
\end{split}
\end{gather}
So the stronger condition \eqref{eq:FTSrank1} means that all three gammas must vanish. Using \eqref{eq:S_i(gamma)} it is then clear that all three local entropies vanish.

Conversely, from \eqref{eq:gamma(S_i)}, $S_A=S_B=S_C=0$ implies that each of the three $\gamma$'s vanish and the rank 1 condition is satisfied. Hence FTS rank 1 is equivalent to the class of separable states as in \autoref{tab:merge}.

\subsubsection{Rank 2 and the class of biseparable states}

 A  non-zero state $\Psi$ is rank 2 or less if and only if $T(\Psi,\Psi,\Psi)=0$. To not be rank 1 there must exist some $\Phi$ such that $3T(\Psi, \Psi, \Phi) + \{\Psi , \Phi\}\Psi\not=0$. It was shown in \autoref{sec:rank1} that this is equivalent to only one non-vanishing $\gamma$ matrix.

Using \eqref{eq:S_i(gamma)} it is clear that the choices $\gamma^A\not=0$ or $\gamma^B\not=0$ or $\gamma^C\not=0$  give $S_A=0, S_{B,C}\neq0$ or $S_B=0, S_{C,A}\neq0$ or $S_C=0, S_{A,B,}\neq0$, respectively. These are precisely the conditions for the biseparable class $A$-$BC$ or $B$-$CA$ or $C$-$AB$ presented in \autoref{tab:conventional}.

Conversely, using \eqref{eq:S_i(gamma)}, \eqref{eq:gamma(S_i)} and the fact that the local entropies and $\tr(\gamma^{\dag}\gamma)$ are positive semidefinite, we find that all states in the biseparable class are rank 2, the particular subdivision being given by the corresponding non-zero $\gamma$. Hence FTS rank 2  is equivalent to the class of biseparable states as in \autoref{tab:merge}.

\subsubsection{Rank 3 and the class of W-states}

A  non-zero state $\Psi$ is rank 3 if $q(\Psi)=-2\Det a=0$ but $T(\Psi, \Psi, \Psi)\neq 0$.
From \eqref{eq:Tofgamma} all three $\gamma$'s are then non-zero but from \eqref{eq:q} all have vanishing determinant.   In this case \eqref{eq:S_i(gamma)} implies that all three local entropies are non-zero but $\Det a=0$.  So all rank 3 $\Psi$ belong to the W-class.

Conversely, from \eqref{eq:S_i(gamma)} it is clear that no two $\gamma$'s may simultaneously vanish when all three $S$'s $>0$. We saw in \autoref{sec:rank1} that $T(\Psi, \Psi, \Psi)=0$ implied at least two of the $\gamma$'s vanish. Consequently, for all W-states $T(\Psi, \Psi, \Psi)\not=0$ and, therefore, all W-states are rank 3.  Hence FTS rank 3 is equivalent to the class of W-states as in \autoref{tab:merge}.

\subsubsection{Rank 4 and the class of GHZ-states}

The rank 4 condition is given by $q(\Psi)\neq0$ and, since for the three-qubit FTS $q(\Psi)=-2 \Det a$, we immediately see that the set of rank 4 states is equivalent to the GHZ class of genuine tripartite entanglement  as in \autoref{tab:merge}.

Note, $\Aut(\mathfrak{M}(\mathfrak{J}_\mathds{C}))$ acts transitively only on rank 4 states with the same value of $q(\Psi)$ as in the standard treatment. The GHZ class really corresponds to a continuous space of orbits parametrised by $q$.

In summary, we have demonstrated that each rank corresponds to one of the entanglement classes described in \autoref{sec:conventional}. The fact that these classes are truly distinct (no overlap)  follows immediately from the manifest invariance of the rank conditions.

\subsection{SLOCC orbits}
\label{sec:orbits}

We now turn our attention to the  coset parametrisation of the entanglement classes. The coset space of each orbit $(i=1,2,3,4)$ is given by $G/H_i$ where $G=[SL(2, \mathds{C})]^3$ is the SLOCC group and $H_i \subset [SL(2, \mathds{C})]^3$ is the stability subgroup leaving the representative state of the $i $th orbit invariant. We proceed by considering the infinitesimal action of $\AutM$ on the representative states of each class. The subalgebra annihilating the representative state gives, upon exponentiation, the stability group $H$.

For  the class of Freudenthal triple systems considered here the Lie algebra $\mathfrak{Aut}(\mathfrak{M(J)})$ is given by
\begin{equation}\label{eq:FTSalgebra}
\mathfrak{Aut(M(J))= J\oplus J\oplus Str(J)},
\end{equation}
where $\mathfrak{Str(J)}$ is the Lie algebra of $\Str(\mathfrak{J})$  given by  $\mathfrak{Str(J)}=L_\mathfrak{J}\oplus \Der(\mathfrak{J})$ \cite{Jacobson:1968, Jacobson:1971}.  $L_\mathfrak{J}$ is the set of left Jordan multiplications by elements in $\mathfrak{J}$, i.e. $L_X(Y)=X\circ Y$ for $X, Y \in \mathfrak{J}$. Its centre is given by scalar multiples of the identity and we may decompose $\mathfrak{Str}(\mathfrak{J})=L_\mathds{1}\mathds{F}\oplus \mathfrak{Str}_0(\mathfrak{J})$. Here, $\mathfrak{Str}_0(\mathfrak{J})$ is the reduced structure group Lie algebra which is given by  $\mathfrak{Str}_0(\mathfrak{J})=L_{\mathfrak{J}'}\oplus \Der(\mathfrak{J})$, where $\mathfrak{J}'$ is the set of traceless Jordan algebra elements.

The Lie algebra action on a generic FTS element $(\alpha, \beta, A, B)$ is given by
\begin{equation}
\begin{split}
\alpha'=&-\alpha \tr C + \Tr (X, B),\\
\beta' =&\phantom{-\ \,}\beta \tr C + \Tr (Y, A),\\
A'     =&\phantom{-\ \,}L_C(A)+D(A)+\beta X + Y\times B,\\
B'     =&-L_C(B)+D(B)+ \alpha Y + X\times A.
\end{split}
\end{equation}
where $L_C\in L_\mathfrak{J}$ and  $D\in \Der(\mathfrak{J})$ come from the action of $\mathfrak{Str}(\mathfrak{J})=L_\mathfrak{J}\oplus \Der(\mathfrak{J})$ \cite{Freudenthal:1954,Koecher:1967,Jacobson:1971,Kantor:1973,Shukuzawa:2006,Gutt:2008}. The product $X\times Y$ is defined in \eqref{eq:FreuProduct}.

Let us now focus on the relevant example for three qubits, $\mathfrak{J}=\mathfrak{J}_\mathds{C}$. In this case $\Der(\mathfrak{J}_\mathds{C})$ is empty due to the associativity of $\mathfrak{J}_\mathds{C}$. Consequently, $\mathfrak{Str}(\mathfrak{J}_\mathds{C})=L_\mathds{1}\mathds{F}\oplus \mathfrak{Str}_0(\mathfrak{J}_\mathds{C})$  has complex dimension 3, while $\mathfrak{Str}_0(\mathfrak{J}_\mathds{C})$ is now simply $L_{\mathfrak{J}'}$ and has complex dimension 2. Recall, $\mathfrak{Str}(\mathfrak{J}_\mathds{C})$   and $\mathfrak{Str}_0(\mathfrak{J}_\mathds{C})$ generate $[SO(2, \mathds{C})]^3$ and $[SO(2, \mathds{C})]^2$,  respectively the structure and reduced structure groups of $\mathfrak{J}_\mathds{C}$. The Lie algebra action transforming a state $(\alpha, \beta, A, B)\to (\alpha', \beta', A', B')$ may now be summarised by:
\begin{equation}
\begin{split}
\alpha'=&-\alpha \tr C + \Tr (X, B),\\
\beta' =&\phantom{-\ \,}\beta \tr C + \Tr (Y, A),\\
A'     =&\phantom{-\ \,}L_C(A)+\beta X + Y\times B,\\
B'     =&-L_C(B)+ \alpha Y + X\times A.
\end{split}
\end{equation}
and we may now determine $G/H_i$.

\subsubsection{Rank 1 and the class of separable states}
\begin{gather}
\begin{split}
\ket{\psi}&=\ket{111}\\
\Leftrightarrow\Psi&=(1, 0, (0,0,0), (0,0,0))
\end{split}\\
\begin{array}{c@{\ =\ }c*{4}{@{\ }c}}
\alpha' & -\tr C & \Rightarrow & \tr C & = & 0,\\
\beta'  & 0,     &             &       &   &   \\
A'      & 0,     &             &       &   &   \\
B'      & Y      & \Rightarrow & Y     & = & 0.
\end{array}
\end{gather}
So $H_{1}$ is parameterised by 5 complex numbers, two of which belong to $L_{C'}\in L_{\mathfrak{J}'}=\mathfrak{Str}_0(\mathfrak{J}_\mathds{C})$ and so generate $[SO(2, \mathds{C})]^2$. The remaining three complex parameters from $X\in\mathfrak{J}_\mathds{C}$ generate translations. Hence, denoting semi-direct product by $\ltimes$,
\begin{equation}
\frac{G}{H_1}=\frac{[SL(2,\mathds{C})]^3}{[SO(2,\mathds{C})]^2\ltimes \mathds{C}^3}.
\end{equation}
with complex dimension 4.

\subsubsection{Rank 2 and the class of biseparable states}
\begin{gather}
\begin{split}
\ket{\psi}&=\ket{111}+\ket{001}\\
\Leftrightarrow\Psi&=(1, 0, (1,0,0), (0,0,0))
\end{split}\\
\begin{array}{ccccc@{\ =\ }l}
\alpha' & = & -\tr C      & \Rightarrow & \tr C & 0,      \\
\beta'  & = & \Tr(Y, A)   & \Rightarrow & Y_1   & 0,      \\
A'      & = & L_{C}(A)    & \Rightarrow & C_1   & 0,      \\
B'      & = & Y+X\times A & \Rightarrow & Y_1   & Y_2+X_3 \\
        &   &             &             &       & Y_3+X_2 \\
        &   &             &             &       & 0,
\end{array}
\end{gather}
where $X=(X_1, X_2, X_3)$, $Y=(Y_1, Y_2, Y_3)$ and we have used $X\times A=(X_2A_3+A_2X_3, X_1A_3+A_1X_3, X_1A_2+A_1X_2)$. So $H_{2}$ is parameterised by 4 complex numbers. Three parameters, the one of $L_C$ and two of $Y$, combine to generate  $O(3, \mathds{C})$. The remaining parameter $X_1$, a singlet under the $O(3, \mathds{C})$,  generates a translation.  Hence,
\begin{equation}
\frac{G}{H_2}=\frac{[SL(2,\mathds{C})]^3}{O(3,\mathds{C})\times \mathds{C}}.
\end{equation}
with complex dimension 5.

\subsubsection{Rank 3 and the class of W states}
\begin{gather}
\begin{split}
\ket{\psi}&=\ket{010}+\ket{001}+\ket{100}\\
\Leftrightarrow\Psi&=(0, 0, (1,1,1), (0,0,0))
\end{split}\\
\begin{array}{*{6}{c}l}
\alpha' & = & 0,        &             &            &   &    \\
\beta'  & = & \Tr(Y, A) & \Rightarrow & \Tr(Y)     & = & 0, \\
A'      & = & L_{C}(A)  & \Rightarrow & C\circ A   & = & C  \\
        &   &           &             &            & = & 0, \\
B'      & = & X\times A & \Rightarrow & -X+\Tr(X)A & = & 0  \\
        &   &           & \Rightarrow & X          & = & 0,
\end{array}
\end{gather}
where  we have used the identity
\begin{equation}
\begin{split}
X\times A&=\phantom{+}X\circ A-\half[\Tr(X)A+\Tr(A)X]\\
         &\phantom{=}+\half[\Tr(X)\Tr(A)+\Tr(X, A)]\mathds{1}.
\end{split}
\end{equation}
See, for example, \cite{Jacobson:1968, Jacobson:1971}. So $H_{3}$ is parameterised by 2 complex numbers, namely the traceless part of $Y$ which generates 2-dimensional translations.  Hence,
\begin{equation}
\frac{G}{H_3}=\frac{[SL(2,\mathds{C})]^3}{\mathds{C}^2}.
\end{equation}
with complex dimension 7.

\subsubsection{Rank 4 and the class of GHZ states}
\begin{gather}
\begin{split}
\ket{\psi}&=\ket{000}+\ket{111}\\
\Leftrightarrow\Psi&=(1, 1, (0, 0, 0), (0,0,0))
\end{split}\\
\begin{array}{c@{\ =\ }c@{\ \Rightarrow\ }c@{\ =\ }c}
\alpha' & -\tr C           & \tr C & 0,\\
\beta'  & \phantom{-}\tr C & \tr C & 0,\\
A'      & \phantom{-}X     & X     & 0,\\
B'      & \phantom{-}Y     & Y     & 0.
\end{array}
\end{gather}
So $H_{4}$ is parameterised by 2 complex numbers, the traceless part of $L_C$, which spans $L_{\mathfrak{J}'}=\mathfrak{Str}_0(\mathfrak{J}_\mathds{C})$ and therefore generates $[SO(2, \mathds{C})]^2$.  Hence,
\begin{equation}
\frac{G}{H_4}=\frac{[SL(2,\mathds{C})]^3}{[SO(2, \mathds{C})]^2}.
\end{equation}
with complex dimension 7. Note, the GHZ class is actually a continuous space of orbits parameterised by one complex number, the quartic norm $q$.

These results are summarised in \autoref{tab:3qubitscosets}.  To be clear, in the preceding analysis we have regarded the three-qubit state as a  point in $\mathds{C}^2\times\mathds{C}^2\times\mathds{C}^2$, the philosophy adopted in, for example, \cite{Linden:1997qd, Carteret:2000-1, Sudbery:2001}. We could have equally well considered the projective Hilbert space regarding states as rays in $\mathds{C}^2\times\mathds{C}^2\times\mathds{C}^2$, that is, identifying states related by a global complex scalar factor, as was done in \cite{Miyake:2002, Miyake:2003, Brody:2007}. The coset spaces obtained in this case are also presented in \autoref{tab:3qubitscosets}, the dimensions of which agree with the results of \cite{Miyake:2002, Gelfand:1994}. Note that the three-qubit separable projective coset  is just a direct product of  three individual qubit cosets
$SL(2,\mathds{C})/SO(2,\mathds{C})\ltimes \mathds{C}$.	Furthermore, the biseparable projective coset is just the direct product of  the two entangled qubits coset $[SL(2,\mathds{C})]^2/O(3,\mathds{C})$ and an individual qubit coset.
\begin{table*}
\caption{Coset spaces of the orbits of the 3-qubit state space $\mathds{C}^2\times\mathds{C}^2\times\mathds{C}^2$ under the action of the SLOCC group $[SL(2, \mathds{C})]^3$.\label{tab:3qubitscosets}}
\begin{ruledtabular}
\begin{tabular}{cccD{c}cD{c}cc}
& Class & FTS Rank & \text{Orbits} & dim &	\text{Projective orbits} & dim & \\
\hline
& Separable   & 1 & \frac{[SL(2,\mathds{C})]^3}{[SO(2,\mathds{C})]^2\ltimes\mathds{C}^3} & 4 & \frac{[SL(2,\mathds{C})]^3}{[SO(2,\mathds{C})\ltimes\mathds{C}]^3}                    & 3 & \\
& Biseparable & 2 & \frac{[SL(2,\mathds{C})]^3}{O(3,\mathds{C})\times\mathds{C}}         & 5 & \frac{[SL(2,\mathds{C})]^3}{O(3,\mathds{C})\times(SO(2,\mathds{C})\ltimes\mathds{C})} & 4 & \\
& W           & 3 & \frac{[SL(2,\mathds{C})]^3}{\mathds{C}^2}                            & 7 & \frac{[SL(2,\mathds{C})]^3}{SO(2,\mathds{C})\ltimes\mathds{C}^2}                      & 6 & \\
& GHZ         & 4 & \frac{[SL(2,\mathds{C})]^3}{[SO(2,\mathds{C})]^2}                    & 7 & \frac{[SL(2,\mathds{C})]^3}{[SO(2,\mathds{C})]^2}                                     & 7 &
\end{tabular}
\end{ruledtabular}
\end{table*}
The case of real qubits is treated in \hyperref[sec:real]{appendix B}.

\section{Conclusions}
\label{sec:conclusions}

We have provided an alternative way of classifying three-qubit entanglement based on the rank of a Freudenthal triple system defined over the Jordan algebra $\mathfrak{J}_{\mathds{C}}=\mathds{C\oplus C\oplus C}$.  Some of the advantages are as follows.
\begin{enumerate}
\item Since $ \Psi$, $T(\Psi, \Psi, \Psi)$, $\gamma_A$, $\gamma_B$, $\gamma_C$ and $q(\Psi)$ are all tensors under $SL(2,\mathds{C}) \times SL(2,\mathds{C}) \times SL(2,\mathds{C})$, the classification of \autoref{tab:merge} is manifestly SLOCC invariant. Contrast this with the conventional classification of \autoref{tab:conventional} which, although SLOCC invariant, is not manifestly so since only $\psi$ and $\Det a$ are tensors. The $S_A$, $S_B$ and $S_C$ are only LOCC invariants.
\item The FTS approach facilitates the computation of the SLOCC cosets of \autoref{tab:3qubitscosets}, which, as far as we are aware, were hitherto unknown.
\item Jordan algebras and the FTS appearing in \autoref{tab:FTSsummary} have previously entered the physics literature through ``magic'' and extended supergravities  \cite{Gunaydin:1983bi, Gunaydin:1983rk, Gunaydin:1984ak}, and their ranks through the classification of the corresponding black hole solutions \cite{Ferrara:1997ci, Ferrara:1997uz,Pioline:2006ni}. Indeed, although it is logically independent of it,  the present work was inspired by the black-hole/qubit correspondence \cite{Duff:2006uz,Kallosh:2006zs,Levay:2006kf,Ferrara:2006em,Duff:2006ue,Levay:2006pt,Duff:2006rf,Pioline:2006ni,Bellucci:2007gb,Duff:2007wa,Bellucci:2007zi,Levay:2007nm,Borsten:2008ur,Ferrara:2008hw,Borsten:2008,levay-2008,Bellucci:2008sv,Levay:2008mi,Borsten:2008wd,Borsten:2009zy}.  The possible role of Jordan algebras and/or FTS in the context of entanglement was already mentioned in some of these discussions \cite{Kallosh:2006zs,Pioline:2006ni,Duff:2006ue,Levay:2006pt,Duff:2006rf,Duff:2007wa,Borsten:2008,levay-2008,Borsten:2008wd}, but we hope the explicit construction of the present paper opens the door to a quantum information interpretation of  the other FTS of  \autoref{tab:FTSsummary} \cite{Borsten:2008wd}. In particular, the $E_7$ FTS, defined over the (split) octonionic Jordan algebra $J_{3}^{\mathds{O}}$, corresponds to the
configuration discussed in \cite{Duff:2006ue,Levay:2006pt,Borsten:2008wd,Borsten:2008}, where it was interpreted as describing a particular tripartite entanglement of seven qubits.

\end{enumerate}

\begin{acknowledgments}

We thank Sergio Ferrara, Peter Levay and Alessio Marrani for useful discussions.  This work was supported in part by the STFC under rolling grant ST/G000743/1 and by the DOE under grant No. DE-FG02-92ER40706.

\end{acknowledgments}

\appendix

\section{Jordan algebras and the Freudenthal triple system}\label{sec:JordanandFTS}

\subsection{Jordan algebras}
\label{sec:Jordan}

Typically an FTS is defined by an underlying  \emph{Jordan algebra}. A Jordan algebra $\mathfrak{J}$ is vector space defined over a ground field $\mathds{F}$ equipped with a bilinear product satisfying
\begin{equation}
\begin{split}
A\circ B &=B\circ A,\\ A^2\circ (A\circ B)&=A\circ (A^2\circ B), \quad\forall\ A, B \in \mathfrak{J}.
\end{split}
\end{equation}

For our purposes the relevant Jordan algebra is an example of the class of \emph{cubic} Jordan algebras. A cubic Jordan algebra comes equipped with a cubic form $N:\mathfrak{J}\to \mathds{F}$, satisfying $N(\lambda A)=\lambda^3N(A), \quad \forall \lambda \in \mathds{F}, A\in \mathfrak{J}$. Additionally, there is an element $c\in\mathfrak{J}$ satisfying $N(c)=1$, referred to as a \emph{base point}. There is a very general prescription for constructing cubic Jordan algebras,  due to Springer \cite{Springer:1962,McCrimmon:1969,McCrimmon:2004}, for which all the properties of the Jordan algebra are essentially determined by the cubic form.  We sketch this construction here, following closely the conventions of \cite{Krutelevich:2004}.

Let $V$ be a vector space, defined over a ground field $\mathds{F}$, equipped with both a cubic norm, $N:V\to \mathds{F}$, satisfying $N(\lambda A)=\lambda^3N(A), \quad \forall \lambda \in \mathds{F}, A\in V$, and a base point $c\in V$ such that $N(c)=1$. If $N(A, B, C)$, referred to as the full \emph{linearisation} of $N$, defined by
\begin{equation}
\begin{gathered}
N(A, B, C):=\\
\begin{aligned}
\ &\tfrac{1}{6}\big[N(A+B+C)\\
              &-N(A+B)-N(A+C)-N(B+C)\\
              &+N(A)+N(B)+N(C)\big]
\end{aligned}
\end{gathered}
\end{equation}
is trilinear then  one may define the following four maps,
\begin{subequations}\label{eq:cubicdefs}
\begin{enumerate}
\item The trace,
    \begin{equation}
    \begin{split}
    \Tr:V&\to\mathds{F}\\
    A    &\mapsto3N(c, c, A),
    \end{split}
    \end{equation}
\item A quadratic map,
    \begin{equation}
    \begin{split}
    S:V&\to\mathds{F}\\
    A  &\mapsto3N(A, A, c),
    \end{split}
    \end{equation}
\item A bilinear map,
    \begin{equation}
    \begin{split}
    S: V\times V&\to\mathds{F}\\
    (A,B)       &\mapsto6N(A, B, c),
    \end{split}
    \end{equation}
\item A trace bilinear form,
    \begin{equation}\label{eq:tracebilinearform}
    \begin{split}
    \Tr:V\times V&\to\mathds{F}\\
    (A,B)        &\mapsto\Tr(A)\Tr(B)-S(A, B).
    \end{split}
    \end{equation}
\end{enumerate}
\end{subequations}

A cubic Jordan algebra $\mathfrak{J}$, with multiplicative identity $\mathds{1}=c$, may be derived from any such vector space if $N$ is \emph{Jordan cubic}, that is:
\begin{enumerate}
\item The trace bilinear form \eqref{eq:tracebilinearform} is non-degenerate.
\item The quadratic adjoint map, $\sharp\colon\mathfrak{J}\to\mathfrak{J}$, uniquely defined by $\Tr(A^\sharp, B) = 3N(A, A, B)$, satisfies
\begin{equation}\label{eq:Jcubic}
    (A^{\sharp})^\sharp=N(A)A, \qquad \forall A\in \mathfrak{J}.
    \end{equation}
\end{enumerate}
The Jordan product is then defined using,
\begin{equation}
A\circ B = \half\big(A\times B+\Tr(A)B+\Tr(B)A-S(A, B)\mathds{1}\big),
\end{equation}
where, $A\times B$ is the linearisation of the quadratic adjoint,
\begin{equation}\label{eq:FreuProduct}
A\times B = (A+B)^\sharp-A^\sharp-B^\sharp.
\end{equation}

Important examples include the sets of $3\times 3$ Hermitian matrices, which we denote as $J_{3}^{\mathds{A}}$,  defined over the four division algebras $\mathds{A}=\mathds{R}, \mathds{C}, \mathds{H}$ or $\mathds{O}$  (or their split signature cousins) with Jordan product $A\circ B = \tfrac{1}{2}(AB+BA)$, where $AB$ is just the conventional matrix product.  See \cite{Jacobson:1968} for a comprehensive account. In addition  there is the infinite sequence of \emph{spin factors} $\mathds{F}\oplus Q_n$, where $Q_n$ is an $n$-dimensional vector space over $\mathds{F}$ \cite{Jacobson:1961, Jacobson:1968, Krutelevich:2004, McCrimmon:1969, Baez:2001dm}. The relevant example with respect to three qubits, which we denote as $\mathfrak{J}_\mathds{C}$, is simply the threefold direct sum of $\mathds{C}$, i.e. $\mathfrak{J}_\mathds{C}=\mathds{C\oplus C\oplus C}$, the details of which are given in \autoref{FTSthreequbits}.

There are three groups of particular importance related to cubic Jordan algebras. The set of automorphisms, $\Aut(\mathfrak{J})$,  is composed of all linear transformations on $\mathfrak{J}$ that preserve the Jordan product,
\begin{equation}\label{eq:JAut}
\begin{split}
A\circ B&=C\\
\Rightarrow\quad g(A)\circ g(B)&= g(C), \quad \forall\ g\in \Aut(\mathfrak{J}).
\end{split}
\end{equation}
The Lie algebra of $\Aut(\mathfrak{J})$ is given by the set of derivations, $\operatorname{Der}(\mathfrak{J})$, that is, all linear maps $D:\mathfrak{J}\to\mathfrak{J}$ satisfying the Leibniz rule,
\begin{equation}\label{eq:JDer}
D(A\circ B)=D(A)\circ B+A\circ D(B).
\end{equation}
For any Jordan algebra all derivations may be written in the form $\sum_i [L_{A_i}, L_{B_i}]$, where $L_A(B)=A\circ B$ is the left multiplication map \cite{Schafer:1966}.

The \emph{structure} group, $\Str(\mathfrak{J})$, is composed of all linear bijections on $\mathfrak{J}$ that leave the cubic norm $N$ invariant up to a fixed scalar factor,
\begin{equation}
N(g(A))=\lambda N(A), \quad \forall\ g\in \Str(\mathfrak{J}).
\end{equation}
Finally, the \emph{reduced structure} group $\Str_0(\mathfrak{J})$ leaves the cubic norm invariant and therefore consists of those elements in $\Str(\mathfrak{J})$ for which $\lambda =1$ \cite{Schafer:1966, Jacobson:1968, Brown:1969}.

\subsection{The Freudenthal triple system}
\label{sec:FTS}

In general, given a cubic Jordan algebra $\mathfrak{J}$ defined over a field $\mathds{F}$, one is able to construct an FTS by defining the vector space $\mathfrak{M(J)}$,
\begin{equation}
\mathfrak{M(J)}=\mathds{F\oplus F}\oplus \mathfrak{J\oplus J}.
\end{equation}
An arbitrary element $x\in \mathfrak{M(J)}$ may be written as a ``$2\times 2$ matrix'',
\begin{equation}
x=\begin{pmatrix}\alpha&A\\B&\beta\end{pmatrix} \quad\text{where} ~\alpha, \beta\in\mathds{F}\quad\text{and}\quad A, B\in \mathfrak{J}.
\end{equation}
The FTS comes equipped with a non-degenerate bilinear antisymmetric quadratic form, a quartic form and a trilinear triple product \cite{Freudenthal:1954,Brown:1969,Faulkner:1971, Ferrar:1972, Krutelevich:2004}:
\begin{subequations}
\begin{enumerate}
\item Quadratic form $ \{x, y\}$: $\mathfrak{M(J)}\times \mathfrak{M(J)}\to \mathds{F}$
    \begin{equation}\label{eq:bilinearform}
    \begin{gathered}
    \{x, y\}=\alpha\delta-\beta\gamma+\Tr(A,D)-\Tr(B,C),\\
    \text{where\qquad}x=\begin{pmatrix}\alpha&A\\B&\beta\end{pmatrix},\qquad y=\begin{pmatrix}\gamma&C\\D&\delta\end{pmatrix}.
    \end{gathered}
    \end{equation}
\item Quartic form $q:\mathfrak{M(J)}\to \mathds{F}$
    \begin{equation}\label{eq:quarticnorm}
    \begin{split}
    q(x)=&-2[\alpha\beta-\Tr(A, B)]^2 \\
         &-8[\alpha N(A)+\beta N(B)-\Tr(A^\sharp, B^\sharp)].
    \end{split}
    \end{equation}
\item Triple product $T:\mathfrak{M(J)\times M(J)\times M(J)\to M(J)}$ which is uniquely defined by
	\begin{equation}\label{eq:tripleproduct}
	\{T(x, y, w), z\}=q(x, y, w, z)
	\end{equation}
	where $q(x, y, w, z)$ is the full linearisation of $q(x)$ such that $q(x, x, x, x)=q(x)$.
\end{enumerate}
\end{subequations}
Note that all the necessary definitions, such as the cubic and trace bilinear forms, are inherited from the underlying Jordan algebra $\mathfrak{J}$.

\section{\texorpdfstring{The real case $\mathfrak{J}_\mathds{R}=\mathds{R\oplus R\oplus R}$}{The real case J(R) = R + R + R}}
\label{sec:real}

As noted in \cite{Acin:2000,Levay:2006kf}, the case of real qubits or ``rebits'' is qualitatively different from the complex case. An interesting observation is that on restricting to real states the GHZ class actually has two distinct orbits, characterised by the sign of $q(\Psi)$. This difference shows up in the cosets in the different possible real forms of $[SO(2, \mathds{C})]^2$. For positive $q(\Psi)$ there are two disconnected orbits, both with $[SL(2,\mathds{R})]^3/[U(1)]^2$ cosets, while for negative $q(\Psi)$ there is one orbit $[SL(2,\mathds{R})]^3/[SO(1,1, \mathds{R})]^2$. In which of the two positive   $q(\Psi)$ orbits a given state lies is determined by the sign of the eigenvalues of the three $\gamma$'s, as shown in \autoref{tab:realcosets}. This phenomenon also has its counterpart in the black-hole context \cite{Ferrara:1997ci, Ferrara:1997uz,Bellucci:2006xz, Bellucci:2007gb, Bellucci:2008sv, Borsten:2008wd}, where the two disconnected $q(\Psi)>0$ orbits are given by $1/2$-BPS black holes and non-BPS black holes with vanishing central charge respectively \cite{Bellucci:2006xz}.
\begin{table}
\caption{Coset spaces of the orbits of the real case $\mathfrak{J}_\mathds{R}=\mathds{R\oplus R\oplus R}$ under $[SL(2, \mathds{R})]^3$.\label{tab:realcosets}}
\begin{ruledtabular}
\begin{tabular}{cccM{c}D{c}cc}
& Class       & FTS Rank & q(\Psi) & \text{Orbits}                                               & dim & \\
\hline
& Separable   & 1        & =0      & \frac{[SL(2,\mathds{R})]^3}{[SO(1,1)]^2\ltimes\mathds{R}^3} & 4   & \\
& Biseparable & 2        & =0      & \frac{[SL(2,\mathds{R})]^3}{O(2,1)\times\mathds{R}}         & 5   & \\
& W           & 3        & =0      & \frac{[SL(2,\mathds{R})]^3}{\mathds{R}^2}                   & 7   & \\
& GHZ         & 4        & <0      & \frac{[SL(2,\mathds{R})]^3}{[SO(1,1)]^2}                    & 7   & \\
& GHZ         & 4        & >0      & \frac{[SL(2,\mathds{R})]^3}{[U(1)]^2}                       & 7   & \\
& GHZ         & 4        & >0      & \frac{[SL(2,\mathds{R})]^3}{[U(1)]^2}                       & 7   & \\
\end{tabular}
\end{ruledtabular}
\end{table}

\end{document}